\documentclass[useAMS,usenatbib]{mn2e}
\usepackage{graphicx}

\newcommand{\apj}{ApJ}
\newcommand{\apjl}{ApJL}
\newcommand{\aap}{A\&A}
\newcommand{\aaps}{A\&AS}
\newcommand{\aj}{AJ}
\newcommand{\mnras}{MNRAS}

\newcommand{\pasp}{PASP}
\newcommand{\apjs}{ApJS}
\newcommand{\MC}{\multicolumn}

\newcommand{\St}{StWr\,2-21}
\newcommand{\Bo}{BoBn\,1}
\DeclareRobustCommand{\ion}[2]{%
\relax\ifmmode
\ifx\testbx\f
{\mathrm{#1\,\textsc{#2}}}\else
{\mathrm{#1\,\mathsc{#2}}}\fi
\else\textup{#1\,{\mdseries\textsc{#2}}}%
\fi}

\title[The metallicity extremes of the Sagittarius dSph using SALT 
spectroscopy]
{The metallicity extremes of the  Sagittarius dSph:
SALT spectroscopy of PNe\footnotemark[0]\thanks{%
Based on observations obtained with South African Large Telescope (SALT).}
}
\author[A. Y. Kniazev et al.]{%
Alexei Y. Kniazev,$^{1,2}$\thanks{E-mail: akniazev@saao.ac.za (AYK);
a.zijlstra@manchester.ac.uk (AAZ)
}
Albert A. Zijlstra,$^{3,1}$
Eva K. Grebel,$^{4}$
Leonid S. Pilyugin,$^{5}$
\newauthor
Simon Pustilnik,$^{2}$
Petri V\"ais\"anen,$^{1}$
David Buckley,$^{1}$
Yas Hashimoto,$^{1}$
Nicola Loaring,$^{1}$
\newauthor
Encarni Romero,$^{1}$
Martin Still,$^{1}$
Eric B. Burgh,$^{6}$
Kenneth Nordsieck$^{6}$\\
\rule{-4pt}{20pt}
$^{1}$South African Astronomical Observatory, PO Box 9, 7935, Cape Town, 
South Africa\\
$^{2}$Special Astrophysical Observatory, Nizhnij Arkhyz, Karachai-Circassia, 
369167, Russia \\
$^{3}$University of Manchester, School of Physics \& Astronomy, PO Box 88, 
Manchester M60 1QD \\
$^{4}$Astronomisches Rechen-Institut, Zentrum f\"ur Astronomie Heidelberg,
University of Heidelberg, \\ M\"onchhofstr.\ 12--14, D-69120 Heidelberg, 
Germany \\
$^{5}$Main Astronomical Observatory of National Academy of Sciences of Ukraine,
27 Zabolotnogo str., 03680 Kiev, Ukraine \\
$^{6}$Space Astronomy Laboratory, University of Wisconsin, Madison, WI 53706,
USA
}

\begin{document}

\date{Accepted 2007 July    ??. Received 2007 June ??; in original form 20?? October ??}

\pagerange{\pageref{firstpage}--\pageref{lastpage}} \pubyear{2008}

\maketitle

\label{firstpage}

\begin{abstract}
In this work we present the first spectroscopic results obtained with the
Southern African Large Telescope (SALT) telescope during its
perfomance-verification phase. We find that the Sagittarius dwarf spheroidal
galaxy (Sgr) Sgr contains a youngest stellar population with $\rm
[O/H]\approx-0.2$ and age $t>1\,\rm Gyr$, and an oldest population with
[O/H]\,$=-2.0$. The values are based on spectra of two planetary nebulae
(PNe), using empirical abundance determinations.  We calculated abundances
for O, N, Ne, Ar, S, Cl, Fe, C and He. We confirm the high abundances of PN
\St\ with 12+log(O/H) = 8.57$\pm$0.02 dex.  The other PN studied, \Bo, is an
extraordinary object in that the neon abundance exceeds that of oxygen.  The
abundances of S, Ar and Cl in \Bo\ yield the original stellar metallicity,
corresponding to 12+log(O/H) = 6.72$\pm$0.16 dex which is 1/110 of the
solar value. The actual [O/H] is much higher: third dredge-up enriched the
material by a factor of $\sim$12 in oxygen, $\sim$240 in nitrogen and
$\sim$70 in neon. Neon as well as nitrogen and
oxygen content may have been produced in the intershell of low-mass
AGB stars. Well defined broad WR lines are present in the
spectrum of \St\ and absent in the spectrum of \Bo. This puts the
fraction of [WR]-type central PNe stars to 67\% for dSph galaxies.
\end{abstract}


\begin{table*}
\caption{Observational details of target PNe}
\label{t:Obs}
\begin{tabular}{lccccccc} \hline
Object    &  Right Ascension & Declination   & Date        & Exp.time     & Spectral Range & Slit    & V$_\odot$     \\
Name      &            \MC{2}{c}{J2000}      &             &  (sec)       &       (\AA)    &(arcsec) & (km s$^{-1}$) \\ \hline
\Bo       &  00 37 16.03     & $-$13 42 58.5 & 10.10.2006  & 2$\times$600 & 3500--6630     &  1.0    & 181$\pm$4     \\
	  &                  &               & 13.10.2006  & 3$\times$600 & 6000--9030     &  1.5    & 187$\pm$9     \\
\St       &  19 14 23.35     & $-$32 34 16.6 & 26.10.2006  & 2$\times$600 & 3500--6630     &  1.0    & 131$\pm$3     \\ \hline
\end{tabular}
\end{table*}

\begin{figure*}
 \includegraphics[clip=,angle=-90,width=15.0cm]{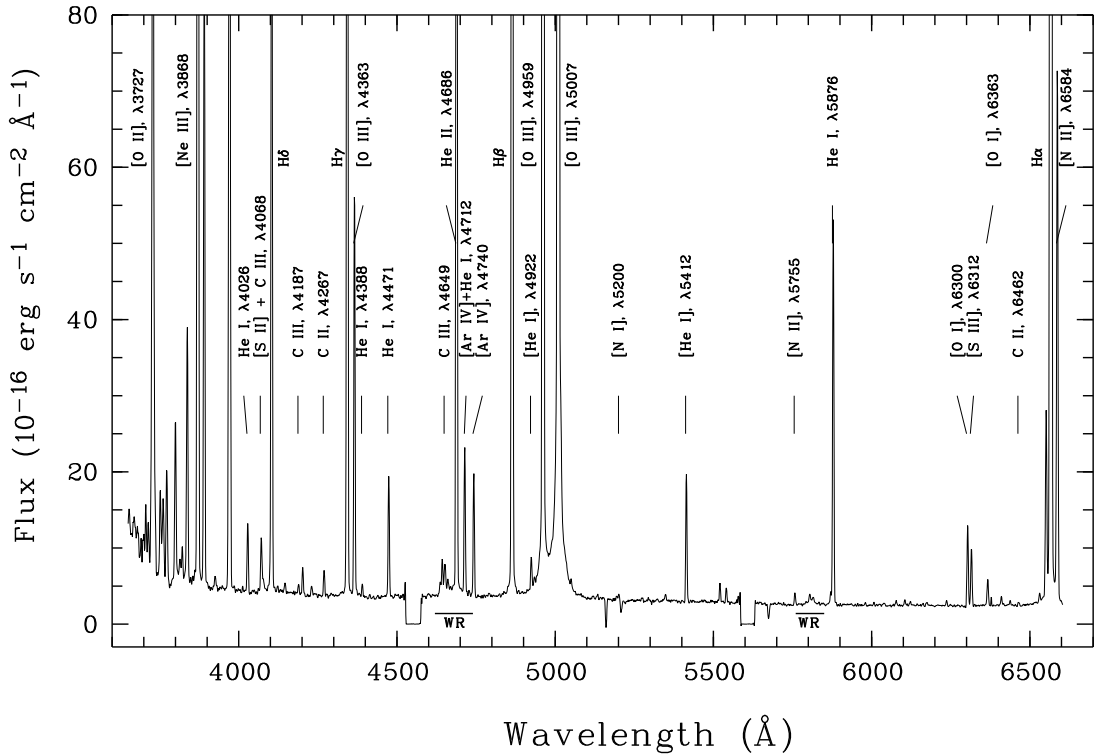}
  \caption{
One-dimensional reduced spectrum of the planetary nebula \St.
The spectrum covers the wavelength range 3500--6630 \AA.
Most of the detected stronger emission lines are marked.
All detected lines are listed in Table~\ref{t:Intens}.
The locations of detected stellar WR lines  are also indicated.
   \label{fig:PNe_spec_StWr}}
\end{figure*}

\section[]{Introduction}

The most common morphological type of the dwarf galaxies in the Local
Group (LG) are the dwarf spheroidals (dSphs). They are also the least
massive, least luminous and most gas-deficient galaxies in the LG.
The dSph galaxies are mostly found as satellites of larger galaxies,
and their properties are likely affected by their dominant
neighbors. The main characteristics are a small size, a lack of
interstellar gas and young stars, and a range of metallicities that
extend to comparatively high values for their luminosity.  The causes
of these properties are disputed \citep*{Gre03}. Gas-stripping by ram
pressure is likely to be involved, and the steep increase of
metallicity with age could be caused by removal of the primordial gas
reservoir and/or accretion of metal-rich gas from the large neighbour
\citep[e.g.,][]{ZGWPHM06}.

In gas-deficient galaxies, accurate nebular abundances can still be
obtained from spectra of planetary nebulae (PNe). These provide
information on elements that are not easily observed in stellar
absorption-line spectra.  For the Local Group and other nearby
galaxies these abundance data can also be combined with star formation
histories derived from color-magnitude diagrams of resolved stars and
ideally with stellar spectroscopic metallicities \citep[e.g.,][ and
references therein]{Ketal07a,Ketal07b}.  In turn, this yields deeper
insights into galaxy evolution, in particular on the overall chemical
evolution of galaxies as a function of time.  Only in two dSph
galaxies in the LG have PNe been detected and observed to date: two in
Fornax \citep{Danz78,L08,Fornax} and four in Sagittarius
\citep{ZW96,ZGWPHM06}.

With a heliocentric distance of $\sim 26$ kpc to its central region
\citep{Mon04}, the Sagittarius dwarf is the closest known dSph.  Sgr
was discovered by \citet*{IGI94,IGI95}.  It is strongly disrupted by
its interaction with the Milky Way
\citep{DP01,HW01,N03,M03,P04}. Since Sgr is located behind the rich
stellar population of the Galactic bulge, studies of the stellar
population in this galaxy are difficult.  \citet{Ma98} found that
individual stars in Sgr have metallicities in the range $\rm
-1.58\le[Fe/H]\le-0.71$. Evidence for a more metal-rich population was
presented by \citet{Bon04}, who derived $\rm [Fe/H] \sim -0.25$ based
on spectroscopy of 10 red giant stars, while \citet{SBBMMZ07} have
found even higher metallicities between [Fe/H] = $-$0.9 and 0.  The
existence of a significant, much more metal-poor population in Sgr is
shown by the metal-poor clusters M\,54, Arp\,2 and Ter\,8
\citep{LS97,LS00}.

\citet{ZW96} discovered two PNe in the Sagittarius dwarf spheroidal
galaxy that were previously cataloged as Galactic PNe: Wray~16$-$423
and He~2$-$436.  The first spectroscopy for them was published by
\citet{W97} and a detailed analysis based on ground-based spectra and
radio-continuum data was carried out by \citet{D00}. Two further PNe
were analyzed in \citet{ZGWPHM06}: a Sgr-core member, \St, and a
possible leading-tail member, \Bo.  The first two PNe showed identical
abundances, of [Fe/H]$=-0.55\pm0.05$ \citep{W97,D00}.  The abundances
of the two recent objects are less well defined, being based on less
accurate spectroscopy. However, current results indicate that one
shows a very high abundance for a dwarf galaxy ([Fe/H]$\approx-0.25$),
while the other, in contrast, has one of the lowest abundances of any
known PN ([Fe/H]$\approx-2$). \Bo\ is located towards a tidal tail
\citep{ZGWPHM06}, but its association with Sgr is not as secure as that
of the other three; it could also be an unrelated Galactic halo
object.

The goal of our present work is to improve our knowledge of the
elemental abundances of the last two PNe in the Sagittarius dSph
through new high-quality spectra obtained with the new effectively
8-meter diameter Southern African Large Telescope (SALT).  The
contents of this paper are organized as follows:
\S~\ref{txt:Obs_and_Red} gives the description of all observations and
data reduction.  In \S~\ref{txt:results} we present our results, and
discuss them in \S~\ref{txt:disc}.  The conclusions drawn from this
study are summarized in \S~\ref{txt:summ}.

\begin{figure*}
 \includegraphics[clip=,angle=-90,width=15.0cm]{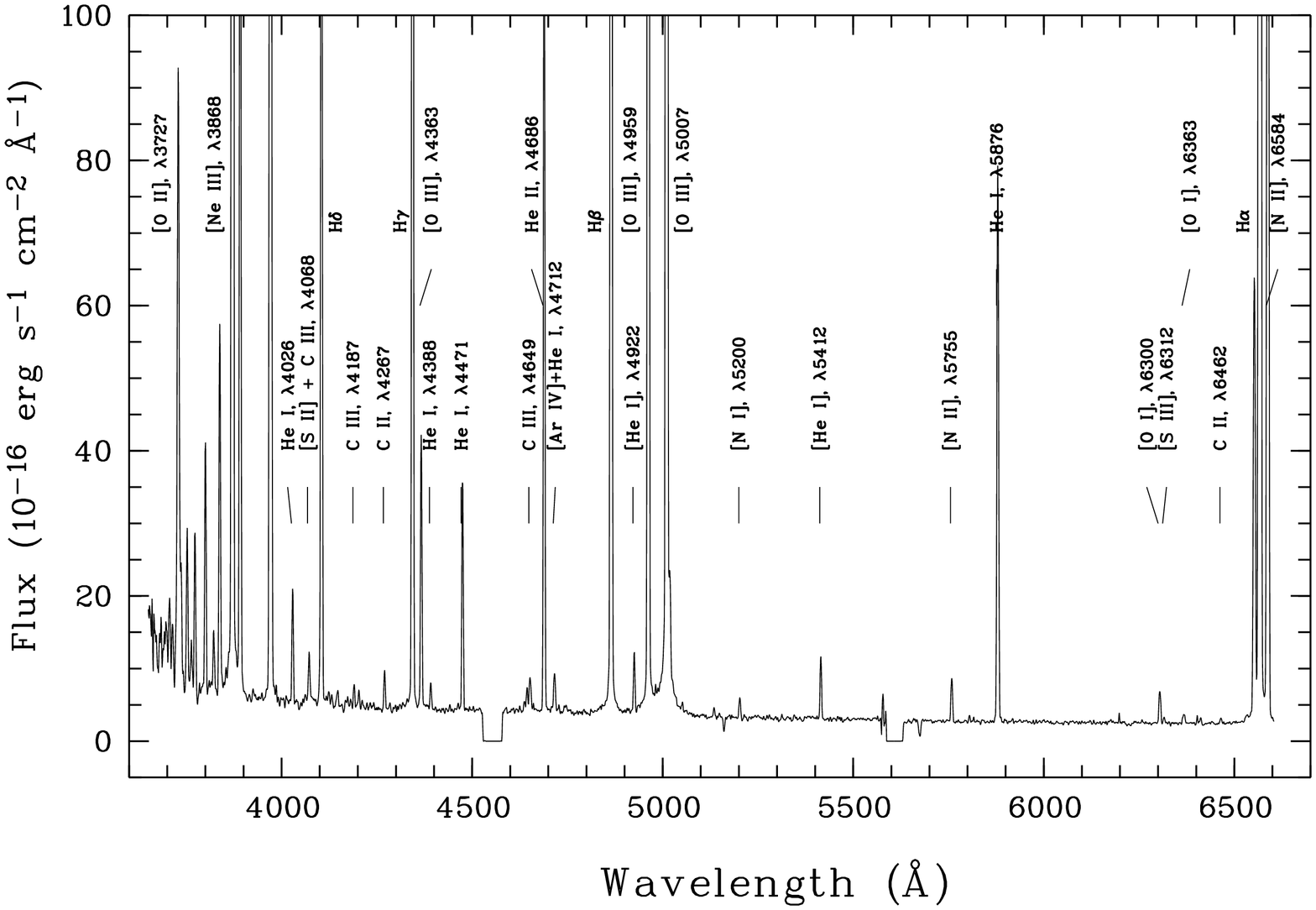}
 \includegraphics[clip=,angle=-90,width=15.0cm]{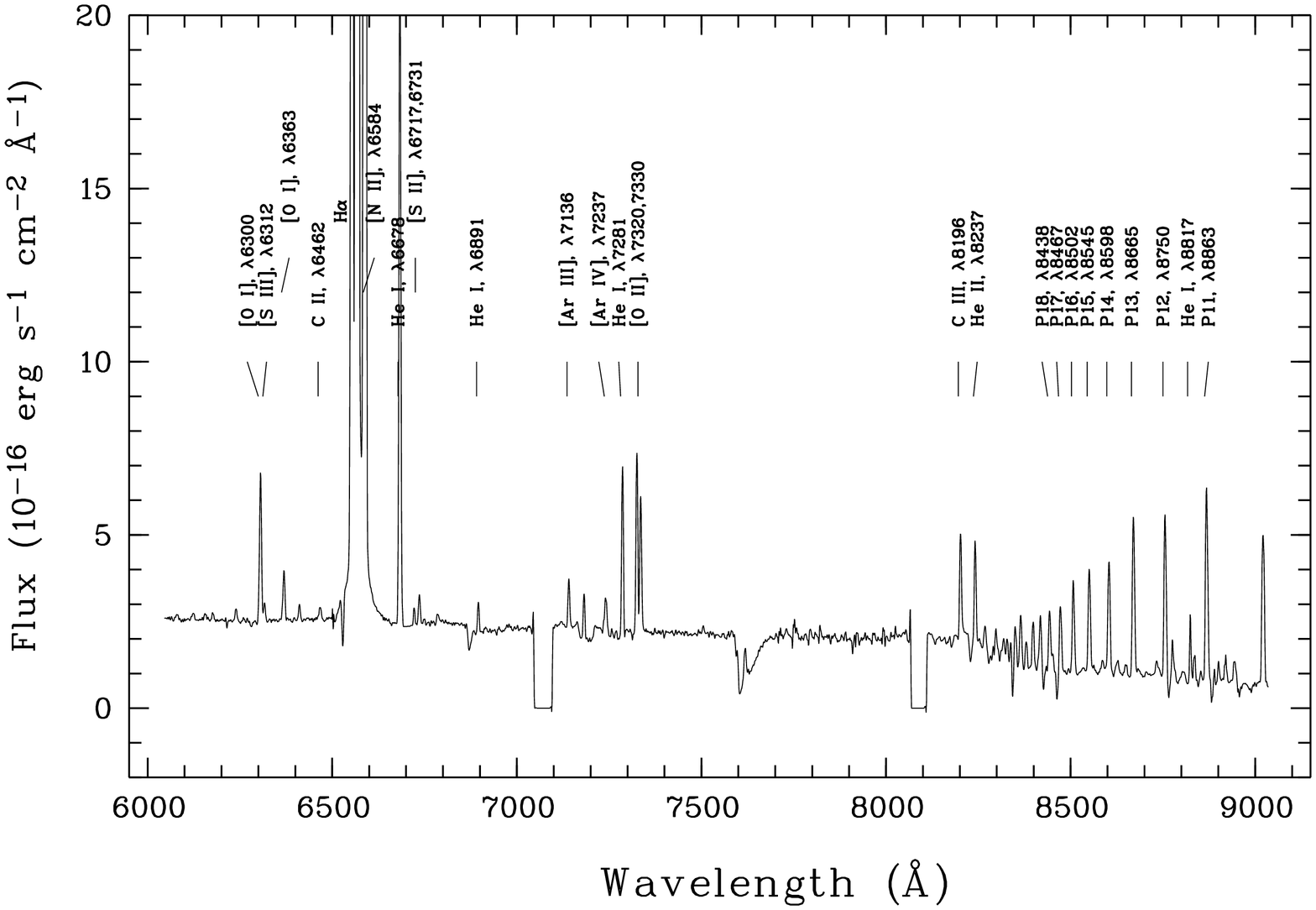}
  \caption{
The one-dimensional reduced spectra of the planetary nebula \Bo.
The top spectrum covers a wavelength range of 3500--6630 \AA\ and
bottom spectrum covers a wavelength range of 6000--9050 \AA.
Most of the detected strong emission lines are marked.
All detected lines are listed in Table~\ref{t:Intens}.
   \label{fig:PNe_spec_BoBn}}
\end{figure*}

\begin{table*}
\centering{
\caption{Line intensities of the studied PNe (part I)}
\label{t:Intens}
\begin{tabular}{lcccc} \hline
\rule{0pt}{10pt}
& \MC{2}{c}{\St}  & \MC{2}{c}{\Bo}   \\ \hline
\rule{0pt}{10pt}
$\lambda_{0}$(\AA) Ion                  & F($\lambda$)/F(H$\beta$)&I($\lambda$)/I(H$\beta$) & F($\lambda$)/F(H$\beta$)&I($\lambda$)/I(H$\beta$) \\ \hline
3727\ [O\ {\sc ii}]\                    & 0.3434$\pm$0.0119  & 0.3584$\pm$0.0132  & 0.1638$\pm$0.0092 & 0.1644$\pm$0.0095 \\
3750\ H12\                              & ---                & ---                & 0.0343$\pm$0.0020 & 0.0345$\pm$0.0021 \\
3771\ H11\                              & 0.0340$\pm$0.0016  & 0.0354$\pm$0.0017  & 0.0336$\pm$0.0023 & 0.0338$\pm$0.0023 \\
3798\ H10\                              & 0.0500$\pm$0.0023  & 0.0520$\pm$0.0025  & 0.0530$\pm$0.0029 & 0.0532$\pm$0.0030 \\
3819\ He\ {\sc i}\                      & 0.0095$\pm$0.0015  & 0.0099$\pm$0.0015  & 0.0157$\pm$0.0015 & 0.0158$\pm$0.0016 \\
3835\ H9\                               & 0.0728$\pm$0.0029  & 0.0756$\pm$0.0032  & 0.0766$\pm$0.0031 & 0.0768$\pm$0.0033 \\
3868\ [Ne\ {\sc iii}]\                  & 0.7317$\pm$0.0236  & 0.7590$\pm$0.0258  & 2.0668$\pm$0.0672 & 2.0730$\pm$0.0710 \\
3889\ He\ {\sc i}\ +\ H8\               & 0.1760$\pm$0.0061  & 0.1824$\pm$0.0066  & 0.2388$\pm$0.0087 & 0.2395$\pm$0.0091 \\
3923\ He\ {\sc ii}\                     & 0.0041$\pm$0.0008  & 0.0042$\pm$0.0008  & ---               & ---               \\
3967\ [Ne\ {\sc iii}]\ +\ H7\           & 0.3416$\pm$0.0108  & 0.3529$\pm$0.0117  & 0.7299$\pm$0.0224 & 0.7319$\pm$0.0236 \\
4026\ He\ {\sc i}\                      & 0.0187$\pm$0.0013  & 0.0193$\pm$0.0013  & 0.0244$\pm$0.0012 & 0.0245$\pm$0.0013 \\
4068\ [S\ {\sc ii}]\ +\ C\ {\sc iii}\   & 0.0162$\pm$0.0015  & 0.0166$\pm$0.0016  & 0.0128$\pm$0.0012 & 0.0129$\pm$0.0012 \\
4076\ [S\ {\sc ii}]\                    & 0.0040$\pm$0.0014  & 0.0042$\pm$0.0014  & ---               & ---               \\
4101\ H$\delta$\                        & 0.2483$\pm$0.0081  & 0.2552$\pm$0.0086  & 0.2622$\pm$0.0085 & 0.2628$\pm$0.0088 \\
4121\ He\ {\sc i}\                      & 0.0020$\pm$0.0010  & 0.0020$\pm$0.0010  & ---               & ---               \\
4144\ He\ {\sc i}\                      & 0.0031$\pm$0.0011  & 0.0032$\pm$0.0011  & 0.0035$\pm$0.0007 & 0.0035$\pm$0.0007 \\
4187\ C\ {\sc iii}\                     & 0.0030$\pm$0.0011  & 0.0030$\pm$0.0011  & 0.0042$\pm$0.0007 & 0.0042$\pm$0.0007 \\
4200\ He\ {\sc ii}\ +N\  {\sc iii}\     & 0.0072$\pm$0.0010  & 0.0074$\pm$0.0011  & 0.0028$\pm$0.0005 & 0.0028$\pm$0.0005 \\
4227\ [Fe\ {\sc v}]\                    & 0.0032$\pm$0.0012  & 0.0033$\pm$0.0013  & ---               & ---               \\
4267\ C\ {\sc ii}\                      & 0.0072$\pm$0.0011  & 0.0074$\pm$0.0011  & 0.0076$\pm$0.0006 & 0.0076$\pm$0.0006 \\
4340\ H$\gamma$\                        & 0.4415$\pm$0.0144  & 0.4496$\pm$0.0149  & 0.4699$\pm$0.0153 & 0.4706$\pm$0.0156 \\
4363\ [O\ {\sc iii}]\                   & 0.1087$\pm$0.0039  & 0.1105$\pm$0.0040  & 0.0561$\pm$0.0024 & 0.0561$\pm$0.0024 \\
4388\ He\ {\sc i}\                      & 0.0039$\pm$0.0010  & 0.0039$\pm$0.0011  & 0.0057$\pm$0.0007 & 0.0058$\pm$0.0007 \\
4438\ He\ {\sc i}\                      & ---                & ---                & 0.0009$\pm$0.0003 & 0.0009$\pm$0.0003 \\
4471\ He\ {\sc i}\                      & 0.0332$\pm$0.0017  & 0.0336$\pm$0.0017  & 0.0460$\pm$0.0020 & 0.0461$\pm$0.0020 \\
4634\ N\  {\sc iii}\                    & 0.0039$\pm$0.0022  & 0.0039$\pm$0.0022  & 0.0021$\pm$0.0014 & 0.0021$\pm$0.0014 \\
4640\ N\  {\sc iii}\                    & 0.0100$\pm$0.0023  & 0.0101$\pm$0.0023  & 0.0050$\pm$0.0014 & 0.0050$\pm$0.0014 \\
4649\ C\ {\sc iii}\                     & 0.0115$\pm$0.0033  & 0.0115$\pm$0.0034  & 0.0094$\pm$0.0021 & 0.0094$\pm$0.0021 \\
4658\ [Fe\ {\sc iii}]\                  & 0.0053$\pm$0.0032  & 0.0053$\pm$0.0032  & 0.0013$\pm$0.0008 & 0.0013$\pm$0.0008 \\
4686\ He\ {\sc ii}\                     & 0.4177$\pm$0.0127  & 0.4202$\pm$0.0128  & 0.1738$\pm$0.0055 & 0.1739$\pm$0.0055 \\
4712\ [Ar\ {\sc iv]}\ +\ He\ {\sc i}\   & 0.0397$\pm$0.0026  & 0.0399$\pm$0.0026  & 0.0087$\pm$0.0014 & 0.0087$\pm$0.0014 \\
4740\ [Ar\ {\sc iv]}\                   & 0.0326$\pm$0.0025  & 0.0328$\pm$0.0025  & 0.0017$\pm$0.0014 & 0.0017$\pm$0.0014 \\
4861\ H$\beta$\                         & 1.0000$\pm$0.0309  & 1.0000$\pm$0.0309  & 1.0000$\pm$0.0315 & 1.0000$\pm$0.0315 \\
4922\ He\ {\sc i}\                      & 0.0133$\pm$0.0018  & 0.0132$\pm$0.0018  & 0.0117$\pm$0.0008 & 0.0117$\pm$0.0008 \\
4959\ [O\ {\sc iii}]\                   & 3.5942$\pm$0.1217  & 3.5829$\pm$0.1214  & 1.1543$\pm$0.0390 & 1.1540$\pm$0.0391 \\
5007\ [O\ {\sc iii}]\                   &10.3838$\pm$0.3345~~&10.3353$\pm$0.3335~~& 3.4256$\pm$0.1102 & 3.4242$\pm$0.1103 \\
5048\ He\ {\sc i}\                      & ---                & ---                & 0.0016$\pm$0.0003 & 0.0016$\pm$0.0003 \\
5131\ O\ {\sc i}\                       & ---                & ---                & 0.0021$\pm$0.0005 & 0.0021$\pm$0.0005 \\
5200\ [N\ {\sc i}]\                     & 0.0018$\pm$0.0012  & 0.0018$\pm$0.0012  & 0.0045$\pm$0.0006 & 0.0045$\pm$0.0006 \\
5412\ He\ {\sc ii}\                     & 0.0329$\pm$0.0015  & 0.0324$\pm$0.0015  & 0.0131$\pm$0.0008 & 0.0130$\pm$0.0008 \\
5518\ [Cl\ {\sc iii}]\                  & 0.0045$\pm$0.0009  & 0.0044$\pm$0.0008  & 0.0001$\pm$0.0001 & 0.0001$\pm$0.0001 \\
5538\ [Cl\ {\sc iii}]\                  & 0.0037$\pm$0.0010  & 0.0036$\pm$0.0010  & 0.0002$\pm$0.0001 & 0.0002$\pm$0.0001 \\
5755\ [N\ {\sc ii}]\                    & 0.0032$\pm$0.0012  & 0.0031$\pm$0.0012  & 0.0103$\pm$0.0007 & 0.0103$\pm$0.0007 \\
5801\ C\ {\sc iv}\                      & 0.0021$\pm$0.0012  & 0.0020$\pm$0.0012  & ---               & ---               \\
5812\ C\ {\sc iv}\                      & 0.0012$\pm$0.0011  & 0.0012$\pm$0.0011  & 0.0006$\pm$0.0002 & 0.0006$\pm$0.0002 \\
5869\ He\ {\sc ii}\                     & 0.0035$\pm$0.0012  & 0.0034$\pm$0.0012  & ---               & ---               \\
5876\ He\ {\sc i}\                      & 0.1027$\pm$0.0034  & 0.0997$\pm$0.0034  & 0.1318$\pm$0.0044 & 0.1314$\pm$0.0045 \\
6074\ He\ {\sc ii}\                     & 0.0010$\pm$0.0005  & 0.0009$\pm$0.0005  & ---               & ---               \\
6102\ [K\ {\sc iv}]\                    & 0.0014$\pm$0.0008  & 0.0013$\pm$0.0008  & ---               & ---               \\
6118\ He\ {\sc ii}\                     & 0.0007$\pm$0.0004  & 0.0007$\pm$0.0004  & ---               & ---               \\
6234\ He\ {\sc ii}\                     & 0.0013$\pm$0.0008  & 0.0012$\pm$0.0008  & ---               & ---               \\
6300\ [O\ {\sc i}]\                     & 0.0225$\pm$0.0015  & 0.0216$\pm$0.0014  & 0.0085$\pm$0.0004 & 0.0085$\pm$0.0004 \\
6312\ [S\ {\sc iii}]\                   & 0.0161$\pm$0.0012  & 0.0154$\pm$0.0012  & 0.0013$\pm$0.0002 & 0.0013$\pm$0.0002 \\
6364\ [O\ {\sc i}]\                     & 0.0077$\pm$0.0012  & 0.0074$\pm$0.0012  & 0.0030$\pm$0.0008 & 0.0030$\pm$0.0008 \\
6406\ He\ {\sc ii}\                     & 0.0024$\pm$0.0011  & 0.0023$\pm$0.0010  & 0.0009$\pm$0.0003 & 0.0009$\pm$0.0003 \\
6435\ [Ar\ {\sc v]}\                    & 0.0011$\pm$0.0007  & 0.0011$\pm$0.0007  & ---               & ---               \\
6462\ C\ {\sc ii}\                      & 0.0008$\pm$0.0009  & 0.0007$\pm$0.0009  & 0.0010$\pm$0.0004 & 0.0010$\pm$0.0004 \\
6527\ He\ {\sc ii}\                     & 0.0028$\pm$0.0010  & 0.0026$\pm$0.0010  & ---               & ---               \\
\hline
\end{tabular}
 }
\end{table*}
\begin{table*}
\centering{
\caption{Line intensities of the studied PNe (part II)}
\label{t:Intens1}
\begin{tabular}{lcccc} \hline
\rule{0pt}{10pt}
& \MC{2}{c}{\St}  & \MC{2}{c}{\Bo}   \\ \hline
\rule{0pt}{10pt}
$\lambda_{0}$(\AA) Ion             & F($\lambda$)/F(H$\beta$)&I($\lambda$)/I(H$\beta$) & F($\lambda$)/F(H$\beta$)&I($\lambda$)/I(H$\beta$) \\ \hline
6548\ [N\ {\sc ii}]\               & 0.0609$\pm$0.0165      & 0.0581$\pm$0.0158& 0.1378$\pm$0.0041 & 0.1373$\pm$0.0045 \\
6563\ H$\alpha$\                   & 2.9783$\pm$0.0902      & 2.8448$\pm$0.0936& 2.7786$\pm$0.0839 & 2.7680$\pm$0.0908 \\
6584\ [N\ {\sc ii}]\               & 0.1651$\pm$0.0197      & 0.1577$\pm$0.0189& 0.4134$\pm$0.0166 & 0.4118$\pm$0.0174 \\
6678\ He\ {\sc i}\                 &~0.0265$\pm$0.0012$^a$  & 0.0252$\pm$0.0012& 0.0396$\pm$0.0012 & 0.0394$\pm$0.0013 \\
6717\ [S\ {\sc ii}]\               &~0.0200$\pm$0.0009$^a$  & 0.0190$\pm$0.0009& 0.0009$\pm$0.0001 & 0.0009$\pm$0.0001 \\
6731\ [S\ {\sc ii}]\               &~0.0297$\pm$0.0014$^a$  & 0.0283$\pm$0.0014& 0.0017$\pm$0.0001 & 0.0016$\pm$0.0001 \\
6891\ He\ {\sc ii}\                & ---                    & ---              & 0.0015$\pm$0.0005 & 0.0015$\pm$0.0005 \\
7065\ He\ {\sc i}\                 &~0.0233$\pm$0.0017$^a$  & 0.0220$\pm$0.0017& ---               & ---               \\
7136\ [Ar\ {\sc iii}]\             &~0.0936$\pm$0.0069$^a$  & 0.0884$\pm$0.0066& 0.0026$\pm$0.0006 & 0.0025$\pm$0.0005 \\
7178\ He\ {\sc ii}\                & ---                    & ---              & 0.0023$\pm$0.0006 & 0.0023$\pm$0.0006 \\
7237\ [Ar\ {\sc iv}]\              & ---                    & ---              & 0.0025$\pm$0.0004 & 0.0025$\pm$0.0004 \\
7281\ He\ {\sc i}\                 & ---                    & ---              & 0.0093$\pm$0.0007 & 0.0092$\pm$0.0008 \\
7320\ [O\ {\sc ii}]\               & ---                    & ---              & 0.0104$\pm$0.0004 & 0.0103$\pm$0.0004 \\
7330\ [O\ {\sc ii}]\               & ---                    & ---              & 0.0078$\pm$0.0004 & 0.0077$\pm$0.0004 \\
7500\ He\ {\sc i}\                 & ---                    & ---              & 0.0005$\pm$0.0005 & 0.0005$\pm$0.0005 \\
8196\ C\ {\sc iii}\                & ---                    & ---              & 0.0063$\pm$0.0006 & 0.0062$\pm$0.0006 \\
8237\ He\ {\sc ii}\                & ---                    & ---              & 0.0057$\pm$0.0007 & 0.0056$\pm$0.0007 \\
8359\ P22\ +\ C\ {\sc iii}\        & ---                    & ---              & 0.0032$\pm$0.0006 & 0.0032$\pm$0.0006 \\
8374\ P21\                         & ---                    & ---              & 0.0017$\pm$0.0006 & 0.0016$\pm$0.0006 \\
8392\ P20\                         & ---                    & ---              & 0.0030$\pm$0.0007 & 0.0030$\pm$0.0007 \\
8413\ P19\                         & ---                    & ---              & 0.0034$\pm$0.0005 & 0.0033$\pm$0.0005 \\
8437\ P18\                         & ---                    & ---              & 0.0041$\pm$0.0004 & 0.0041$\pm$0.0004 \\
8467\ P17\                         & ---                    & ---              & 0.0044$\pm$0.0009 & 0.0044$\pm$0.0009 \\
8502\ P16\                         & ---                    & ---              & 0.0060$\pm$0.0008 & 0.0060$\pm$0.0009 \\
8545\ P15\                         & ---                    & ---              & 0.0073$\pm$0.0009 & 0.0072$\pm$0.0009 \\
8598\ P14\                         & ---                    & ---              & 0.0080$\pm$0.0009 & 0.0079$\pm$0.0009 \\
8665\ P13\                         & ---                    & ---              & 0.0112$\pm$0.0009 & 0.0111$\pm$0.0010 \\
8750\ P12\                         & ---                    & ---              & 0.0110$\pm$0.0009 & 0.0109$\pm$0.0010 \\
8817\ He\ {\sc i}\                 & ---                    & ---              & 0.0029$\pm$0.0005 & 0.0029$\pm$0.0005 \\
8863\ P11\                         & ---                    & ---              & 0.0142$\pm$0.0011 & 0.0141$\pm$0.0011 \\
& & & & \\
C(H$\beta$)\ dex          & \MC {2}{c}{0.06$\pm$0.04} & \MC {2}{c}{0.01$\pm$0.04} \\
EW(H$\beta$)\ \AA\        & \MC {2}{c}{ 671$\pm$15}   & \MC {2}{c}{ 900$\pm$20}   \\
\hline
\MC{5}{l}{~~} \\
\MC{5}{l}{$^a$ intensity was taken from \citet{ZGWPHM06}.}\\
\end{tabular}
 }
\end{table*}

\section[]{Observations and Data Reduction}
\label{txt:Obs_and_Red}

\subsection[]{SALT Spectroscopic Observations}

The observations of \St\ and \Bo\ were obtained during the Performance
Verification phase (PV) of the SALT telescope \citep{Buck06,Dono06},
and used the Robert Stobie Spectrograph
\citep[RSS;][]{Burgh03,Kobul03}.  The long-slit spectroscopy mode of
the RSS was used, with three mosaiced 2048$\times$4096 CCD detectors.
The RSS pixel scale is 0\farcs129 and the effective field of view is
8\arcmin\ in diameter.  We utilized a binning factor of 2, to give a
final spatial sampling of 0\farcs258 pixel$^{-1}$.  The Volume Phase
Holographic (VPH) grating GR900 was used in two spectral ranges:
3500--6600 \AA\ and 5900--8930 \AA\ with a final reciprocal dispersion
of $\sim$0.95 \AA\ pixel$^{-1}$ and spectral resolution FWHM of 5--6
\AA.  Each exposure was observed with the spectrograph slit aligned to
the parallactic angle to avoid loss of light due to atmospheric
differential refraction.  As shown in Table~\ref{t:Obs}, each exposure
was broken up into 2--3 sub-exposures, 10 minutes each, to allow for
removal of cosmic rays.  Spectra of ThAr and Xe comparison arcs were
obtained to calibrate the wavelength scale.  Four spectrophotometric
standard stars G~93-48, EG~21, BPM~16274 and SA95-42
\citep{SB83,BS84,Mas88,Oke90} were observed at the parallactic angles
for relative flux calibration.

\subsection[]{Data Reduction}
\label{txt:Red}

The data for each CCD detector were bias and overscan subtracted, gain
corrected, trimmed and cross-talk corrected. After that they were
mosaiced.  The primary data reduction was done using the
IRAF\footnote{IRAF: the Image Reduction and Analysis Facility is
distributed by the National Optical Astronomy Observatory, which is
operated by the Association of Universities for Research in Astronomy,
In. (AURA) under cooperative agreement with the National Science
Foundation (NSF).} package {\it salt}\footnote{ See
http://www.salt.ac.za/partners-login/partners/data-analysis-software/
for more information.}  developed for SALT data.  Cosmic ray removal
was done with the task FILTER/COSMIC task in MIDAS.\footnote{MIDAS is
an acronym for the European Southern Observatory package -- Munich
Image Data Analysis System.}  We used the IRAF software tasks in {\it
twodspec} package to perform the wavelength calibration and to correct
each frame for distortion and tilt. To derive the sensitivity curve,
we fitted the observed spectral energy distribution of the standard
stars by a low-order polynomial.  All sensitivity curves observed
during each night were compared and we found the final curves to have
a precision better than 2--3\% over the whole optical range, except
for the region blueward of $\lambda$3700 where precision decreases to
10--15\%.  Spectra were corrected for the sensitivity curve using the
Sutherland extinction curve.  One-dimensional (1D) spectra were then
extracted using the IRAF APALL task.  All one-dimensional spectra
obtained with the same setup for the same object were then
averaged. Finally, the blue and red parts of the total spectrum of
\Bo\ were combined.  The resulting reduced spectra of \St\ and \Bo\
are shown in Figure~\ref{fig:PNe_spec_StWr} and
Figure~\ref{fig:PNe_spec_BoBn}, respectively.

All emission lines were measured applying the MIDAS programs described in 
detail in \citet{K00, SHOC}:
(1) the software is based on the MIDAS Command Language;
(2) the continuum noise estimation was done using
the absolute median deviation (AMD) estimator;
(3) the continuum was determined with the algorithm from \citet*{Sh_Kn_Li_96};
(4) the programs dealing with the fitting
of emission/absorption line parameters are based on the MIDAS
{\tt FIT} package;
(5) every line was fitted with the Corrected Gauss-Newton method
as a single Gaussian superimposed on the continuum-subtracted spectrum.
All overlapping lines were fitted
simultaneously as a blend of two or more Gaussian features:
the [\ion{O}{i}] $\lambda$6300 and [\ion{S}{iii}] $\lambda$6312,
the H$\alpha$ $\lambda$6563 and [\ion{N}{ii}] $\lambda\lambda$6548,6584
lines, the [\ion{S}{ii}] $\lambda\lambda$6716,6731 lines, and the
[\ion{O}{ii}] $\lambda\lambda$7320,7330 lines.
(6) the final errors in the line intensities, $\sigma_{\rm tot}$, include
two components: $\sigma_{\rm p}$, due to the Poisson statistics of line photon flux,
and $\sigma_{\rm c}$, the error resulting from the creation of
the underlying continuum and calculated using the AMD estimator;
(7) all uncertainties were then propagated in the
calculation of errors in electron number densities, electron temperatures
and element abundances.

For the Wolf-Rayet features the Gaussian decomposition to the narrow lines
and the broad components was also done using the same method.

SALT is a telescope with a variable pupil, where the illuminating beam
changes continuously during the observations. This means that absolute
flux calibration is not possible even using spectrophotometric
standard stars. To calibrate absolute fluxes we used H$\beta$ fluxes
from other sources.  For the H$\beta$ flux calibration of \Bo\ we used
the mean value 3.57$\times$10$^{-13}$ erg cm$^{-2}$ s$^{-1}$
calculated from \citet{KH96} and \citet*{WCP05} which are in close
agreement to each other.  The value for the H$\beta$ flux from
\citet*{PTR91} (2.19$\times$10$^{-13}$ erg cm$^{-2}$ s$^{-1}$) is
about 1.5 times weaker compared to the previous references and was
ignored for this reason.  For the H$\beta$ flux calibration of \St\ we
used the mean value 1.15$\times$10$^{-13}$ erg cm$^{-2}$ s$^{-1}$ from
\citet{ZGWPHM06}.  The resulting spectra in
Figure~\ref{fig:PNe_spec_StWr} and Figure~\ref{fig:PNe_spec_BoBn} are
shown after this calibration was performed.

\begin{table}
\centering{
\caption{Important line ratios, calculated temperatures and electron densities in studied PNe}
\label{t:T_D}
\begin{tabular}{lcc} \hline
\rule{0pt}{10pt}
Value                                                        & \St                   &  \Bo             \\ \hline
$[$\ion{O}{iii}$]$ ($\lambda$4959+$\lambda$5007)/$\lambda$4363 & 127.88$\pm$5.73       & 81.72$\pm$4.04     \\
$[$\ion{N}{ii}$]$  ($\lambda$6548+$\lambda$6584)/$\lambda$5755 & ~69.12$\pm$5.18       & 53.28$\pm$2.06     \\
& & \\
	  $[$\ion{S}{ii}$]$~$\lambda$6731/$\lambda$6717        & 0.674$\pm$0.045       & 0.520$\pm$0.056     \\
	  $[$\ion{Cl}{iii}$]$~$\lambda$5518/$\lambda$5538      & 1.222$\pm$0.414       & 0.594$\pm$1.548     \\
	   $[$\ion{Ar}{iv}$]$~$\lambda$4711/$\lambda$4740      &~1.116$\pm$0.109$^a$   & ---                 \\
& & \\
$T_{\rm e}$(\ion{O}{iii})(K)\                                & 11700$\pm$190         & 13720$\pm$870~      \\
$T_{\rm e}$(\ion{N}{ii})(K)\                                 & 11360$\pm$2200        & 11320$\pm$1630      \\
& & \\
%
$T_{\rm e}$(\ion{Ar}{iii})(K)\                               & 12050$\pm$340         & 13250$\pm$1330      \\
$T_{\rm e}$(\ion{Cl}{iii})(K)\                               & 12050$\pm$340         & 13250$\pm$1330      \\
& & \\
$N_{\rm e}$(\ion{S}{ii} $\lambda$6731/$\lambda$6717)         & 2700$_{-570}^{+850}$  & 9600$_{-4360}^{+27100}$ \\
$N_{\rm e}$(\ion{Cl}{iii} $\lambda$5518/$\lambda$5538)       & 930$_{-930}^{+4450}$  & 13400:$^a$         \\
$N_{\rm e}$(\ion{Ar}{iv}  $\lambda$4711/$\lambda$4740)$^b$   & 2920$_{-1220}^{+1540}$& ---                 \\
\hline
\MC{3}{l}{~~} \\
\MC{3}{l}{$^a$ Has large errors.}\\
\MC{3}{l}{$^b$ Corrected for the \ion{He}{i} $\lambda$4713 line.}\\
\end{tabular}
 }
\end{table}

\begin{table}
\centering{
\caption{Elemental abundances in studied PNe}
\label{t:Chem}
\begin{tabular}{lcc} \hline
\rule{0pt}{10pt}
Value                                                       & \St                  &  \Bo             \\ \hline
O$^{+}$/H$^{+}$($\times$10$^5$)\                            & ~1.14$\pm$0.87       & 0.88$\pm$0.93       \\
O$^{++}$/H$^{+}$($\times$10$^5$)\                           & 22.85$\pm$1.26       & 4.76$\pm$0.83       \\
O$^{+++}$/H$^{+}$($\times$10$^5$)\                          & 12.75$\pm$1.09       & 0.78$\pm$0.18       \\
O/H($\times$10$^5$)\                                        & 36.75$\pm$1.88       & 6.42$\pm$1.26       \\
12+log(O/H)\                                                & ~8.57$\pm$0.02       & 7.81$\pm$0.09       \\
& & \\
N$^{+}$/H$^{+}$($\times$10$^7$)\                            & 23.06$\pm$11.13      & 63.48$\pm$24.38     \\
ICF(N)\                                                     & 23.842               & 6.837               \\
N/H($\times$10$^5$)\                                        & 5.50$\pm$2.65        & 4.34$\pm$1.67       \\
12+log(N/H)\                                                & 7.74$\pm$0.21        & 7.64$\pm$0.17       \\
log(N/O)\                                                   & $-$0.83$\pm$0.21     & $-$0.17$\pm$0.18    \\
& & \\
Ne$^{++}$/H$^{+}$($\times$10$^5$)\                          & 4.55$\pm$0.31        & 7.26$\pm$1.47       \\
ICF(Ne)\                                                    & 1.437                & 1.110               \\
Ne/H($\times$10$^5$)\                                       & 6.54$\pm$0.44        & 8.06$\pm$1.64       \\
12+log(Ne/H)\                                               & 7.82$\pm$0.03        & 7.91$\pm$0.09       \\
log(Ne/O)\                                                  & $-$0.75$\pm$0.04     & 0.10$\pm$0.12       \\
& & \\
S$^{+}$/H$^{+}$($\times$10$^7$)\                            & 1.06$\pm$0.46        & 0.09$\pm$0.09       \\
S$^{++}$/H$^{+}$($\times$10$^7$)\                           & 16.67$\pm$2.14       & 0.99$\pm$0.38       \\
ICF(S)\                                                     & 5.538                & 1.324               \\
S/H($\times$10$^7$)\                                        & 98.21$\pm$12.11      & 1.44$\pm$0.52       \\
12+log(S/H)\                                                & 6.99$\pm$0.05        & 5.16$\pm$0.16       \\
log(S/O)\                                                   & $-$1.57$\pm$0.06     & $-$2.65$\pm$0.18    \\
& & \\
Ar$^{++}$/H$^{+}$($\times$10$^7$)\                          & 5.66$\pm$0.54        & 0.15$\pm$0.06       \\
Ar$^{+++}$/H$^{+}$($\times$10$^7$)\                         & 6.76$\pm$0.51        & 0.22$\pm$0.17       \\
ICF(Ar)\                                                    & 1.067                & 1.002               \\
Ar/H($\times$10$^7$)\                                       & 13.25$\pm$0.80       & 0.37$\pm$0.19       \\
12+log(Ar/H)\                                               & 6.12$\pm$0.03        & 4.57$\pm$0.22       \\
log(Ar/O)\                                                  & $-$2.44$\pm$0.03     & $-$3.24$\pm$0.23    \\
& & \\
Fe$^{++}$/H$^{+}$($\times$10$^7$)\                          & 2.18$\pm$1.60        & 0.54$\pm$0.38       \\
ICF(Fe)\                                                    & 35.185               &  9.786              \\
log(Fe/O)\                                                  & $-$1.68$\pm$0.32     & $-$2.09$\pm$0.31    \\
$[$Fe/H$]$\                                                 & $-$0.66$\pm$0.32     & $-$1.82$\pm$0.30    \\
$[$O/Fe$]$\                                                 & ~0.46$\pm$0.32       & ~0.87$\pm$0.30      \\
& & \\
Cl$^{++}$/H$^{+}$($\times$10$^8$)\                          & 2.92$\pm$0.52        & 0.08$\pm$0.04       \\
ICF(Cl)\                                                    & 4.287                & 1.690               \\
Cl/H($\times$10$^8$)\                                       & 12.50$\pm$2.23       & 0.14$\pm$0.07       \\
12+log(Cl/H)\                                               & 5.10$\pm$0.08        & 3.14$\pm$0.23       \\
log(Cl/O)\                                                  & $-$3.47$\pm$0.08~~   & $-$4.66$\pm$0.24~~  \\
& & \\
He$^{+}$/H$^{+}$($\times$10$^2$)\                           & ~6.99$\pm$0.18       &   8.52$\pm$0.40     \\
He$^{++}$/H$^{+}$($\times$10$^2$)\                          & ~3.61$\pm$0.11       &  ~1.53$\pm$0.05     \\
He/H($\times$10$^2$)\                                       & 10.60$\pm$0.21       &  10.05$\pm$0.41     \\
12+log(He/H)\                                               & 11.03$\pm$0.01       &  11.00$\pm$0.02     \\
& \\
C$^{++}$/H$^{+}$($\times$10$^4$)\                           & 7.27$\pm$1.18        & 7.78$\pm$0.67       \\
C$^{+++}$/H$^{+}$($\times$10$^4$)\                          & 5.14$\pm$1.23        & 5.62$\pm$0.48       \\
ICF(C)\                                                     & 1.050                & 1.184               \\
C/H($\times$10$^4$)\                                        & 13.03$\pm$1.78       & 15.86$\pm$0.98      \\
12+log(C/H)\                                                & 9.11$\pm$0.06        & 9.20$\pm$0.03       \\
log(C/O)\                                                   & 0.55$\pm$0.06        & 1.39$\pm$0.09       \\
\hline
\end{tabular}
 }
\end{table}

\subsection{Physical conditions and determination of heavy element abundances}
\label{txt:Phys}

The spectra are interpreted by the technique of plasma diagnostics,
i.e. assuming that all lines are produced in an isothermal gas at uniform
density and ionization level. As a first step, the reddening correction,
electron temperatures and density were calculated. These steps are repeated
several times, until the values converge.

The measured emission line intensities $F(\lambda)$ were corrected for
reddening using the equation
\begin{equation}
\frac{I(\lambda)}{I({\rm H}\beta)} = \frac{F(\lambda)}{F({\rm H}\beta)} \cdot10^{C({\rm H}\beta)f(\lambda)},
\label{eq:CHb1}
\end{equation}
where $C$(H$\beta$) is the extinction coefficient, $I(\lambda)$ is the
intrinsic line flux and $F(\lambda)$ is the observed line flux
corrected for atmospheric extinction.  $f(\lambda)$ is the reddening
function, taken from \citet{Wh58} and normalized at H$\beta$.
\citet*{ITL94} noted that this reddening function can be approximated
over the whole spectral range with an accuracy better than 5\% by
\begin{equation}
f(\lambda)= 3.15854\cdot10^{-1.02109\lambda}-1
\end{equation}
where $\lambda$ is in units of $\mu$m.  The Balmer theoretical ratios
from \citet{Brocklehurst71} were used with equation~\ref{eq:CHb1} in
an iterative way for the intrinsic hydrogen line intensity ratios for
estimated electron temperature.

For collisionally excited lines (CELs), we calculated
ionic and total element abundances for O, N, S, Ne, Ar, Cl and
Fe using equations from \citet{ISMGT06}.
These recent equations are based on sequences of photoionization
models and used the new atomic data of \citet{St05}.

\citet{ISMGT06} find that the electron temperature $\rm T_e$(\ion{O}{iii})
derived from the [\ion{O}{iii}] ($\lambda$4959+$\lambda$5007)/$\lambda$4363
line ratio using old and new emissivities differ by less than 1\% in the
temperature range 5000--20,000 K.  For this reason the electron temperature
$\rm T_e(\ion{O}{iii})$ was derived from an iterative procedure, using the
equation from \citet{Aller84}:
\begin{equation}
\frac{I(4959 + 5007)}{I(4363)} = C_T \Big[\frac{1+a_1 x}{1+a_2 x}\Big] 10^{1.432/t},
\label{eq:oiii}
\end{equation}
where $\rm t = 10^{-4} T_e (\ion{O}{iii})$
and $\rm x = 10^{-4} N_e t^{-0.5}$.
The parameters $C_T$, $a_1$, $a_2$ were calculated using the following
interpolations:
\begin{equation}
C_T = 8.44 - 1.09 t + 0.5 t^2 - 0.08 t^3,
\end{equation}
\begin{equation}
a_1 = 2\cdot10^{-4} + 3.13\cdot10^{-4}t - 1.6\cdot10^{-4}t^2 + 2.67\cdot10^{-5}t^3,
\end{equation}
\begin{equation}
a_2 = 0.0291 + 0.0253t - 0.0128t^2 + 0.00213t^3
\end{equation}
The density $\rm N_e$ was also derived iteratively, using the [\ion{S}{ii}]
$\lambda$6717/$\lambda$6731 line ratio.
The [\ion{O}{iii}] auroral
$\lambda$4363 line was corrected for recombination excitation following the
equation from \citet{Liu00}.

The line [\ion{N}{ii}] $\lambda$5755 was detected in the spectra of
both studied PNe, which allowed us to determine $\rm T_e$(\ion{N}{ii})
directly from the
\mbox{Q$_{\rm N}$ = [\ion{N}{ii}] ($\lambda$6548+$\lambda$6584)/$\lambda$5755 line ratio}.
It is convenient to have an analytic expression linked the electron 
temperature T$_{\rm e}$(\ion{N}{ii}) to the value of the Q$_{\rm N}$ and
the electron density N$_{\rm e}$. 
To establish such a relation we have obtained the five-level-atom solution
for the  N$^+$ ion, with a recent atomic data. 
The Einstein coefficients for spontaneous transitions and the energy levels 
for five low-lying levels were taken from \citet{galavisetal97}. 
The effective cross sections or effective collision strengths for electron 
impact were taken from \citet{hudsonbell05}. The effective cross sections 
are continuous functions of temperatures, tabulated by 
\citet{hudsonbell05} at a fixed temperatures. The actual effective cross 
sections for a given electron temperature are derived from two-order 
polynomial fits of the data from \citet{hudsonbell05} as a function 
of temperature.

We derived a numerical solution of the five-level-atom for a large range 
of values of electron temperature (within the range 5,000 $\div$ 25,000 K) and 
of electron density (within the range 10 $\div$ 10$^6$ cm$^{-3}$). The analytic 
relation was derived as an approximation of those numerical results.
The following simple expression for the approximation was adopted 
(this type of expression was widely used by different authors) 
\begin{equation}
t = \frac{C_0}{\log (Q_{\rm N})  + C_1 +C_2 \, \log (t) + C_3 \, t}
\label{equation:addgen}
\end{equation}
where t = 10$^{-4}$T$_{\rm e}$(N\,{\sc ii}) and coefficients 
C$_i$ can be a function of the electron density.
At N$_{\rm e}$ $\la$ 10$^3$ cm$^{-3}$ the coefficients C$_i$ are, in fact, 
independent of N$_{\rm e}$, 
and the numerical results are well reproduced by the relation  
\begin{equation}
t = \frac{1.111}{\log (Q_{\rm N})  - 0.892 - 0.144 \, \log (t) + 0.023 \, t}.
\label{equation:addlow}
\end{equation}
derived by \citet{pilyugin07} for N$_{\rm e}$ = 100 cm$^{-3}$. 
At N$_{\rm e}$ $\ga$ 10$^3$ cm$^{-3}$ the dependence of the coefficients 
C$_i$ on the electron density are reproduced by the following relations
\begin{equation}
C_0 = 2.56604 - 1.04750 \, w + 0.23621 \, w^2 - 0.01608 \, w^3  ,
\label{equation:addc0}
\end{equation}
\begin{equation}
C_1 = 2.94332 - 2.53546 \,w + 0.48115 \, w^2 - 0.01988 \, w^3   ,
\label{equation:addc1}
\end{equation}
\begin{equation}
C_2 = -9.03905 + 6.89003 \, w - 1.68270 \, w^2 + 0.12347 \, w^3  ,
\label{equation:addc2}
\end{equation}
\begin{equation}
C_3 = 0.70472 - 0.55891 \, w + 0.14310 \, w^2 - 0.01084 \, w^3 ,
\label{equation:addc3}
\end{equation}
where $w$ = log(N$_{\rm e}$). The  values of 
T$_{\rm e}$(N\,{\sc ii}) derived directly from the numerical solution 
and with the approximation differ by less than 3\% 
at N$_{\rm e}$ $\la$ 10$^5$ cm$^{-3}$ and by around around 5\% 
at N$_{\rm e}$ $\sim$ 10$^6$ cm$^{-3}$ and t $\sim$ 2.5.  
It should be emphasized that at N$_{\rm e}$ $\la$ 10$^3$ cm$^{-3}$
Eqs.(\ref{equation:addc0})-(\ref{equation:addc3}) are not workable, and 
Eq.(\ref{equation:addlow}) should be used in this case. 

We thus used $\rm T_e$(\ion{N}{ii}) from Eq.(\ref{equation:addgen})
for the calculation of N$^+$/H$^+$, O$^+$/H$^+$, S$^+$/H$^+$ and Fe$^{2+}$/H$^+$
abundances.
We calculated also $\rm T_e$(\ion{O}{ii}) and $\rm T_e$(\ion{S}{iii})
using approximations from \citet{ISMGT06}.
We used $\rm T_e$(\ion{S}{iii}) for the calculation of S$^{2+}$/H$^+$,
Cl$^{2+}$/H$^+$ and Ar$^{2+}$/H$^+$ abundances and $\rm T_e$(\ion{O}{iii})
from Eq.(\ref{eq:oiii}) for the calculation of O$^{2+}$/H$^+$ and Ne$^{2+}$/H$^+$
abundances.

For \St\ only the [\ion{O}{ii}] $\lambda\lambda$3727,3729 doublet was
used to calculate O$^+$/H$^+$. In case of \Bo, O$^+$/H$^+$ was
calculated as a weighted average of O$^+$/H$^+$ using intensities of
the [\ion{O}{ii}] $\lambda\lambda$3727,3729 doublet as well as
using the [\ion{O}{ii}] $\lambda\lambda$7320,7330
lines. The contribution to the intensities of the [\ion{O}{ii}]
$\lambda\lambda$7320,7330 lines due to recombination was taken into
account following the correction from \citet{Liu00}.

The detection of a strong nebular \ion{He}{ii} $\lambda$4686 emission implies
the presence of a non-negligible amount of O$^{3+}$.  In this case its
abundance is derived from the relation:
\begin{equation}
\frac{\rm O^{3+}}{\rm H^+} = \frac{{\rm He}^{2+}}{{\rm He}^+} \left(\frac{\rm O^+}{\rm H^+} + \frac{\rm O^{2+}}{\rm H^+}\right)
\label{eq:O3+}
\end{equation}
After that, the total oxygen abundance is equal to
\begin{equation}
\frac{\rm O}{\rm H} = \frac{\rm O^+}{\rm H^+} + \frac{\rm O^{2+}}{\rm H^+} + \frac{\rm O^{3+}}{\rm H^+}       \label{eq:O}
\end{equation}

From measured intensities of optical recombination lines (ORLs), ionic
abundances were calculated using the equation:
\begin{equation}
\rm \frac{N(\rm X^{i+})}{N(\rm H^{+})} = \frac{I_{jk}}{I_{\rm H\beta}} \frac{\lambda_{jk}}{\lambda_{\rm H\beta}} \frac{\alpha_{\rm H\beta}}{\alpha_{jk}}
\end{equation}
where $\rm I_{jk}/I_{\rm H\beta}$ is the intensity ratio of the ionic
line to the H$\beta$ line, $\rm \lambda_{jk}/\lambda_{\rm H\beta}$ is the
wavelength ratio of the ionic line to H$\beta$, $\alpha_{\rm H\beta}$
is the effective recombination coefficient for H$\beta$ and
$\alpha_{jk}$ is the effective recombination coefficient for the ionic
ORL.  Ionic abundances derived from optical recombination lines (ORLs)
depend only weakly on the adopted temperature and are essentially
independent of $N_{\rm e}$. A temperature of $\rm T_e$(\ion{O}{iii})
was assumed throughout.  The ionic abundance of observed \ion{C}{ii}
recombination lines was derived using calculated effective
recombination coefficients from \citet{Dav00}, which include both
radiative and dielectronic recombination processes.  The ionic
abundance of observed \ion{C}{iii} recombination lines was derived
using effective recombination coefficients from \citet*{PPB91}.
The effective recombination coefficient for H$\beta$ was taken from
\citet{PPB91} as well.
Dielectronic recombination coefficients for \ion{C}{iii} lines
were taken from \citet{NS84}.
ICF for C were calculated using \citet{KB94}.

Helium was calculated in the manner described in \citet*{ITL97,IT98,IT04}.
The new \citet{Ben02} fits were used to convert \ion{He}{i} emission
line strengths to singly ionized helium $y^+ =$~He$^+$/H$^+$ abundances.

The [\ion{Cl}{iii}] $\lambda$5518/$\lambda$5538 line ratio and [\ion{Ar}{iv}]
$\lambda$4711/$\lambda$4740 line ratio were also used to calculate the density
for \St.
The TEMDEN task contained within the IRAF NEBULAR package was
used in this case.  Since [\ion{Ar}{iv}] $\lambda$4711 is contaminated by
the \ion{He}{i} $\lambda$4713 line, we have subtracted the latter adopting a
\ion{He}{i} $\lambda$4713/$\lambda$4471 recombination ratio \citep{B72} that
could be approximated for $\rm T_e$ varying between 10,000 and 20,000 K with
a linear equation:
\begin{equation}
\frac{\rm \ion{He}{i} \, \lambda 4713}{\rm \ion{He}{i} \, \lambda 4471} = 0.045 + 0.047 \, t
\end{equation}
Unfortunately, these lines are very weak and have large errors in
case of \Bo.

\section{Results}
\label{txt:results}

The measured heliocentric radial velocities for the two studied PNe
are given in Table~\ref{t:Obs}. The two independent measurements for
\Bo\ are consistent with each other and all measured velocities are
consistent within the uncertainties with velocities published in
\citet{ZGWPHM06}.

Tables~\ref{t:Intens} and \ref{t:Intens1} list the relative
intensities of all detected emission lines relative to H$\beta$
(F($\lambda$)/F(H$\beta$)), the ratios corrected for the extinction
(I($\lambda$)/I(H$\beta$)), as well as the derived extinction
coefficient $C$(H$\beta$), and the equivalent width of the H$\beta$
emission line EW(H$\beta$).  $C$(H$\beta$) combines  the internal
extinction in each PN and the foreground extinction in the Milky Way,
however internal extinction tends to be low in all but the youngest PNe.
The electron temperatures $T_{\rm e}$(\ion{O}{iii}), $T_{\rm
e}$(\ion{N}{ii}), the number densities $N_{\rm e}$(\ion{S}{ii}),
$N_{\rm e}$(\ion{Ar}{iv}), $N_{\rm e}$(\ion{Cl}{iii}) are shown
in Table~\ref{t:T_D} together with the line ratios that were used for their
calculations. The ionic
and total element abundances and ICFs for O, N, Ne, S, Ar, Fe, Cl, C
and He are presented in Table~\ref{t:Chem}.

It is desirable to compare densities derived from diagnostics that are
observed in the same spectrum, in order to compare the same volume in
the nebula.  \citet{SK89} examined densities derived for a large
sample of PNe and have found that densities derived from the
[\ion{O}{ii}], [\ion{S}{ii}], [\ion{Ar}{iv}] and [\ion{Cl}{iii}]
doublet ratios agree to within $\sim$30\%.  \citet{KE92} studied a
sample of 57 PNe having densities in the range 500--10000 sm$^{-3}$
and showed that $N_e$(\ion{O}{ii}) and $N_e$(\ion{S}{ii}) are the same
within the errors.  As seen from Table~\ref{t:T_D}, both PNe have
densities that are equal within the formal observational errors and we
may conclude that the nebulae are roughly homogeneous.

The comparison of measured line intensities for \St\ with those
published by \citet{ZGWPHM06} show good agreement, within the cited
errors.  Our RSS spectra have a spectral resolution of 5--6 \AA, which
is about two times better than the spectral resolution of 11 \AA\
\citet{ZGWPHM06} obtained with EFOSC2 on the ESO 3.6m telescope. This
allows for a more accurate detection of faint lines in the same
spectral region and the possibility to fit in detail (see
Figure~\ref{fig:PNe_spec_StWr}) both the blue and the red WR bumps clearly
evident in our spectrum (see Section~\ref{txt:WR} for more details).
There are many weak He lines detected in the spectrum of \St,
especially in the spectral region 5800--6550 \AA.  Since our observed
spectral region for \St\ is limited to $<6630$\AA\ and our detected
intensities for the relatively bright lines are very similar to those
cited in \citet[][ see their Table~2]{ZGWPHM06}, we additionally used
those spectral data (for $\lambda>6630$ \AA) in our calculations of
abundances of S, Ar and He.  All these lines are marked in
Table~\ref{t:Intens1}.  The determined abundances of various elements
using the SALT spectrum are fairly consistent with those published by
\citet{ZGWPHM06} that have been obtained from models.  
\citet{ZGWPHM06} did not determine errors on their abundances.  Our
data also allows us, for the first time, to determine abundances of Fe
and Cl in \St.

\citet{TP79}, \citet{Bar80}, \citet{PTR91} and \citet{WCP05} carried out
previous spectroscopic observations of \Bo\ PN using KPNO 2.1m, CTIO 4-m and
INT 2.5m telescopes.  Our data for \Bo\ are considerably better than the
earlier spectra, in terms of signal-to-noise ratio, resolution and spectral
coverage, yielding a signal-to-noise ratio, for example, of 23 for the
[\ion{O}{iii}] $\lambda$4363 line.  With our new SALT data we securely detect,
for the first time, the weak [\ion{S}{ii}] $\lambda\lambda$6717,6731 lines
(I(6717+6731) $\approx$ 0.0025 I(H$\beta$)) and determine the electron number
density to be three times as large as found before \citep{TP79,PTR91}.
Unfortunately, even with our data the errors for [\ion{S}{ii}]
$\lambda\lambda$6717,6731 lines are still large, yielding large errors for
the density.  This leads to significant errors of practically all calculated
abundances based on CELs.  Our abundances for \Bo\ are consistent within
the uncertainties cited by \citep{PTR91}.  Abundances for \Bo\ have been also
obtained by \citet*{HHM97} and by \citet{ZGWPHM06} from models. However, both
papers are based on the previous observational data \citep[mainly
from][]{PTR91}, where the most important lines for determination of Ar, S and
C were obtained with large uncertainties, the [\ion{N}{ii}] $\lambda5755$ line
was outside of the observed spectral region and [\ion{S}{iii}] $\lambda6312$
was not detected.

The dichotomy between the ORL and CEL abundances is a well known
problem in nebular astrophysics \citep[see][ and references
therein]{Liu03}.  Heavy-element abundances derived from ORLs are
systematically higher than those derived from CELs.  \citet{TP79}
calculated the abundance of C for \Bo\ using the ultraviolet
\ion{C}{iii}] $\lambda$1909, comparing it with the C abundance calculated
from optical recombination lines and found agreement.  Our total C
abundance (see Table~\ref{t:Chem}) derived using the five ORL
\ion{C}{ii} and \ion{C}{iii} lines is consistent, within the quoted
uncertainties, with their value of 12+log(C/H) = 9.16$\pm$0.15.
Unfortunately, there were no ORL lines detected for other elements
and we have to conclude that a higher S/N spectrum will be needed to
identify any discrepancy between the ORL and CEL abundances for both
\St\ and \Bo.

\begin{figure}
{\centering
 \includegraphics[clip=,angle=-90,width=8.0cm]{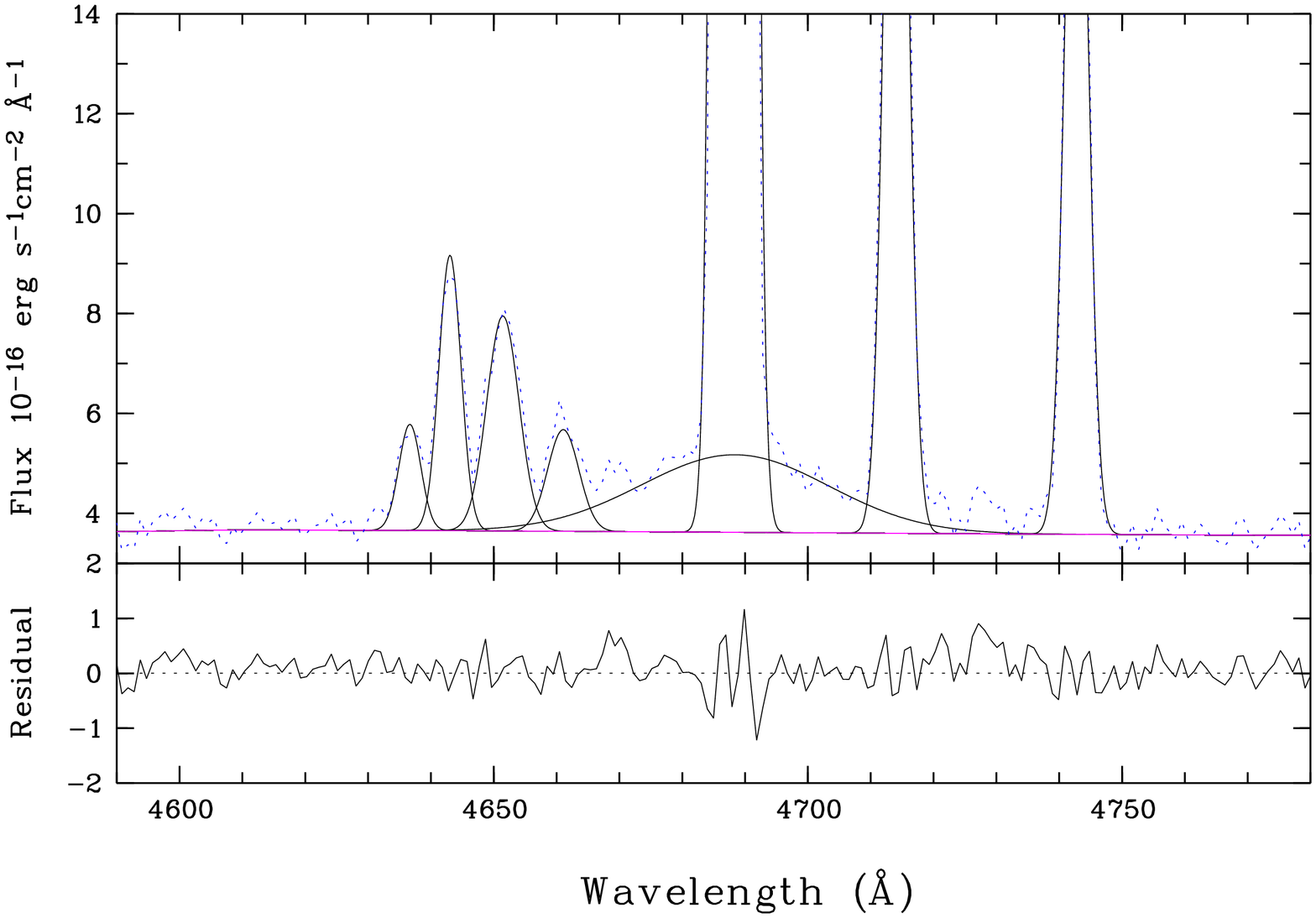}
 \includegraphics[clip=,angle=-90,width=8.0cm]{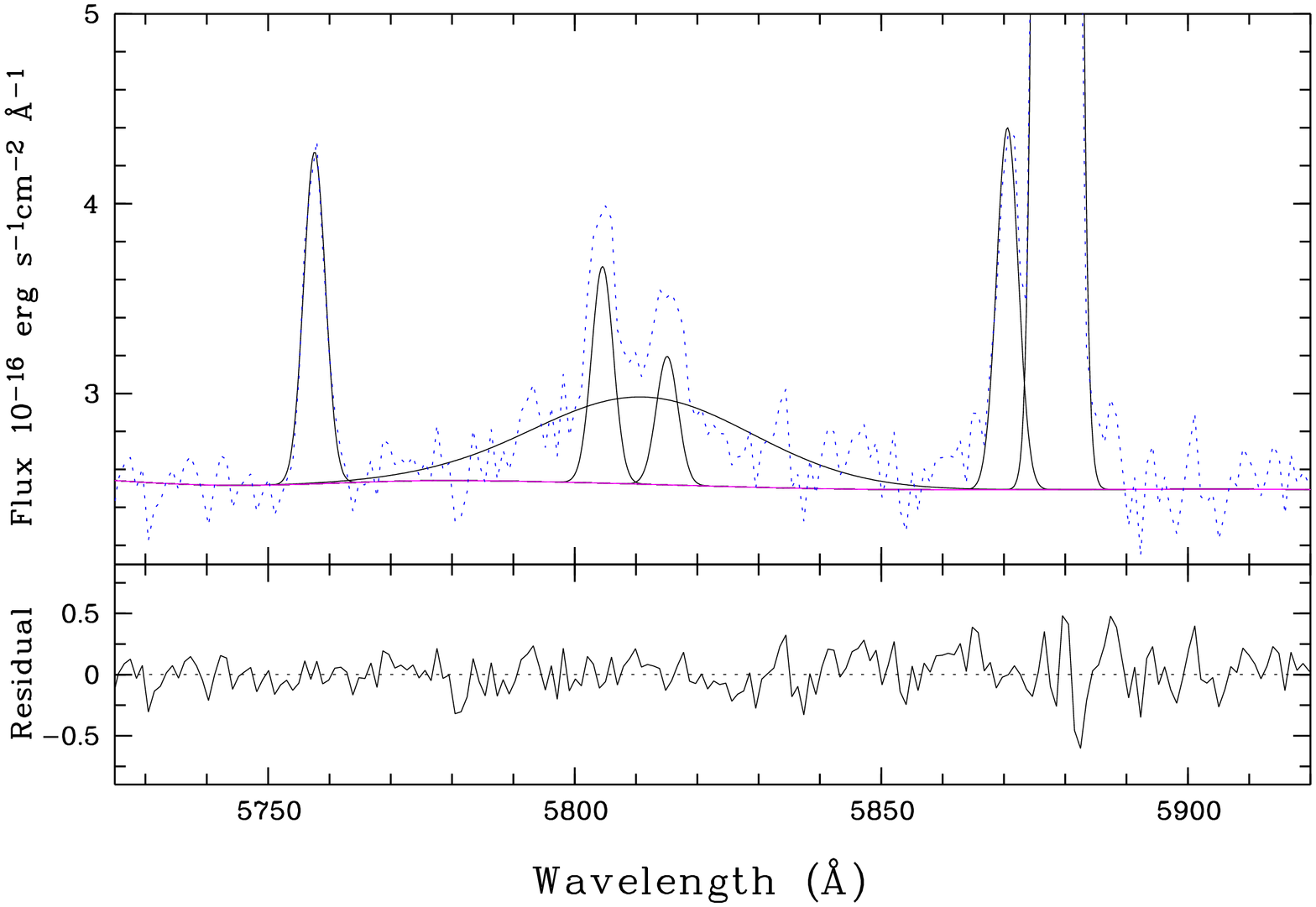}
}
\caption{
Gaussian multicomponent fitting of the blue (top) and the red (bottom)
Wolf-Rayet bumps of \St.
In the top part of each panel
the observed spectrum in the fitted region (dashed lines), fitted gaussians
and created continuum level are shown.
In the bottom part of each panel residuals between the observed spectrum
and the continuum plus fitted lines are plotted.
The following lines from left to right are separated:
{\it Top panel:} The narrow lines (FWHM=4.1 \AA) \ion{N}{iii} $\lambda$4634,
\ion{N}{iii} $\lambda$4640, \ion{C}{iii} $\lambda\lambda$4647+4650,
[\ion{Fe}{iii}] $\lambda$4658, \ion{He}{ii} $\lambda$4686
(narrow and broad components),
narrow [\ion{Ar}{iv}]+\ion{He}{ii} $\lambda\lambda$4711+4713,
and [\ion{Ar}{iv}] $\lambda$4740;
{\it Bottom panel:} The narrow lines (FWHM=5.15 \AA) [\ion{N}{ii}] $\lambda$5755,
\ion{C}{iv} $\lambda$5801 and \ion{C}{iv} $\lambda$5812,
\ion{He}{ii} $\lambda$5869 and \ion{He}{i} $\lambda$5876.
Also the broad Wolf-Rayet bump \ion{C}{iv} was fitted with the center at $\lambda$5805.
}
\label{fig:StWr_WR_fig}
\end{figure}

\subsection{Wolf-Rayet features}
\label{txt:WR}

\citet{ZGWPHM06} reported a detection, based on spectral synthesis, of two
broad Wolf-Rayet (WR) features at 3820 and 5805 \AA in their \St\ spectrum.
There is confirming evidence in our SALT data of broad blue (4640--4750 \AA)
and red (5750--5900 \AA) WR bumps as it is shown on
Figure~\ref{fig:StWr_WR_fig}.  We attempted to confirm the detection of broad
WR features in our \St\ spectrum, and if possible to infer their properties.

The analysis methods used are the same as those described in
Section~\ref{txt:Red}. A local underlying continuum at all the
studied regions  was defined for the total
spectrum using the algorithm from \citet{Sh_Kn_Li_96}.  The
continuum-subtracted spectral region of each studied spectral region
was fitted simultaneously as a blend of a single wide and many narrow
Gaussian features.  Wavelengths of the narrow Gaussian features used
were selected from the list of emission lines located at the
respective spectral regions.  This list was created by the authors and
used by programs described in Section~\ref{txt:Red} for automatic
measuring of emission lines.  The exact wavelengths of the narrow
emission lines were kept fixed during the
fitting procedure.  During this procedure, the FWHMs of most narrow
emission lines were equal to each other, and equal to the FWHM of the
strongest narrow emission line observed in the studied region, namely
\ion{He}{ii} $\lambda$4686 for the blue WR bump spectral region and
\ion{He}{i} $\lambda$5876 for the red WR bump region. The FWHM of the
strongest narrow emission line, the FWHM of the broad component and its
position, and the intensities of all emission lines were used as free
parameters during the fitting procedure.  Any Gaussian feature was
formally detected if the derived intensity was greater than the total
error found for this line (Signal-to-Noise ratio $>$ 1).  Gaussian
features with detected Signal-to-Noise ratios less than unity were
excluded, and the fitting procedure was repeated.

The results from the fitting procedures and residuals are shown in
Figure~\ref{fig:StWr_WR_fig}, while final intensities of all narrow emission
lines are given in Table~\ref{t:Intens}. It seems that the WR wide components
are indeed present and have the following parameters resulting from the
fitting procedure described above: (1) the broad component of the \ion{He}{ii}
$\lambda$4686 line (EW = 16$\pm$5 \AA, FWHM = 35$\pm$15 \AA, flux =
$58\cdot10^{-16}$ ergs\ s$^{-1}$cm$^{-2}$); (2) the broad component of
\ion{C}{iv} centred at $\lambda$5805 (EW = 9 \AA, FWHM = 42$\pm$3 \AA, flux =
$22\cdot10^{-16}$ ergs\ s$^{-1}$cm$^{-2}$).  Our fitting method is too
unstable in the region of the \ion{O}{vi} $\lambda$3822 feature, found
previously by \citet{ZGWPHM06}, to make any definitive conclusions of its
existence in our data.

Both of the broad features which were found, and reported above,
show the same velocity range of $\approx$2200 km s$^{-1}$.
The resulting detected line ratio
\ion{He}{ii} $\lambda$4686/\ion{C}{iv}$\lambda$5805 $\approx$~2.6
supports the conclusion of \citet{ZGWPHM06}
that the central star of \St\ is a hot star midway between
subclasses 2 and 3 \mbox{([WO~2--3])} according to the classification by
\citet{AN03}.  The FWHMs of the detected broad lines also fit well
those shown in \citet{AN03} for \mbox{[WO~2--3]} subclasses.
No broad WR features were detected in the spectrum of \Bo using the described
fitting method.

Finally, four out of six PNe studied in dSphs \citep[][ this
work]{W97,ZGWPHM06,Fornax,L08} show WR features in their spectra. This
fraction (67\%) is much higher than the fraction of [WR]-type central PNe
stars in both the Galactic disk ($\sim$~6.5\%) and the Galactic bulge
\citep[$\sim$~18\%;][]{Go01,Go04}.  This high fraction of [WR]-type of
central PNe stars, as noted by \citet{ZGWPHM06}, supports models
in which the probability of star to develop a WR wind is determined by
parameters of the progenitor star rather than models where this
occurence is a random event caused by a late thermal pulse \citep{Ges07}.

\section{Discussion}
\label{txt:disc}

\subsection{Additional Enrichment in PN Progenitors}
\label{txt:enrich}

\begin{figure}
{\centering
 \includegraphics[clip=,angle=0,width=8.5cm]{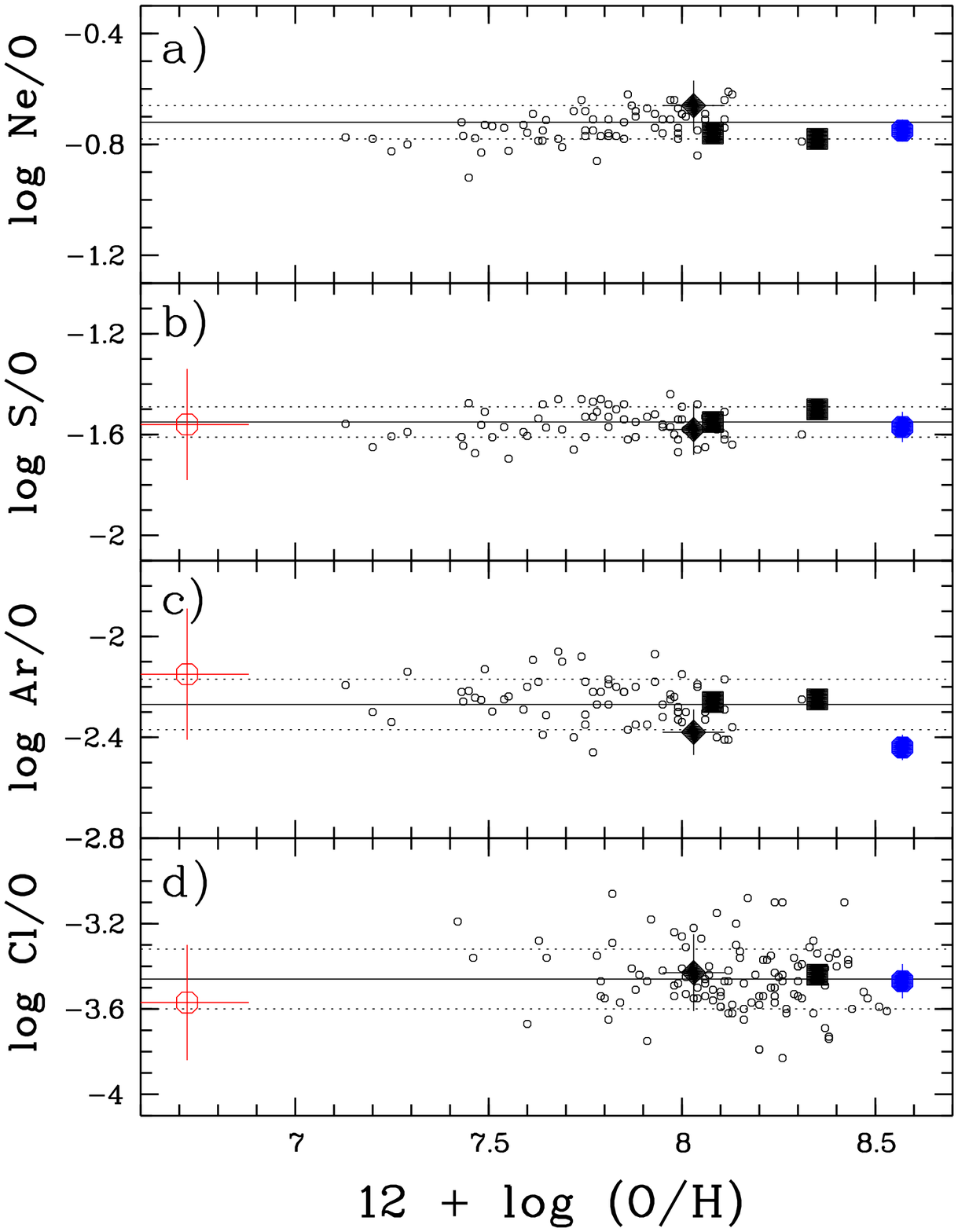}
}
\caption{ $\alpha$-element-to-oxygen abundance ratios for log(Ne/O), log(S/O),
log(Ar/O) and log(Cl/O) for \ion{H}{ii} regions with their 1$\sigma$ errors
(short-dashed lines) as a function of oxygen abundance \citep{IT99,ISMGT06}.
Data from \citet{IT99,Kniazev03,ISMGT06} are overplotted.  Data for PNe in
Fornax (filled diamond) and in the Sagittarius dSph galaxies are shown
with their 1$\sigma$ errors.
Ratios for PN in Fornax are corrected for self-pollution in oxygen by 0.27 dex
\citep{Fornax}.  Ratios shown for \Bo\ in Sgr are corrected for self-pollution
in oxygen by 1.09 dex as discussed in Section~\ref{txt:enrich},
and are plotted as open circles.
Data for \St\ from current work are plotted as filled circles.
Data
for two PNe in the Sagittarius dSph galaxy from \citet{W97} were recalculated
in the same way as described in Section~\ref{txt:Phys} and are plotted as filled
squares.
}
\label{fig:Abun_ratios}
\end{figure}

\ion{H}{ii} region abundances mainly provide information about
$\alpha$-process elements, which are produced predominantly in
short-lived massive stars. Because of their common origin, log(Ne/O),
log(S/O) and log(Ar/O) should be constant and show no dependence on
the oxygen abundance.  \citet{IT99} very accurately measured these
$\alpha$-element-to-oxygen abundance ratios in a large sample of
\ion{H}{ii} regions in blue compact galaxies.  They found that
log(Ne/O) = $-0.72\pm0.06$, log(S/O) = $-1.55\pm0.06$ and log(Ar/O) =
$-2.27\pm0.10$, as shown in Figure~\ref{fig:Abun_ratios}.  Recent
spectrophotometric results of \citet{ISMGT06} for a large sample of
\ion{H}{ii} galaxies from the Sloan Digital Sky Survey DR3 data
\citep{DR3} support this conclusion. In addition, they found no
significant trends with the oxygen abundance for the log(Cl/O) ratio.
Using their published data we calculated weighted mean for the log(Cl/O)
ratio as $-3.46\pm0.14$ and plot these data in the bottom panel of
Figure~\ref{fig:Abun_ratios}.

In contrast to \ion{H}{ii} regions, some elemental abundances in PNe
are affected by the nucleosynthesis in the PN progenitors.  Newly
synthesized material can be dredged up by convection in the envelope,
significantly altering abundances of He, C, and N in the surface
layers during the evolution of the PN progenitor stars on the giant
branch and asymptotic giant branch (AGB). Also a certain amount of
oxygen can be mixed in during the thermally pulsing phase of AGB
evolution \citep{KB94,P00,Leisy2006}.  In combination, it means that
only the Ne, S, Cl and Ar abundances, observed in both H\,{\sc ii}
regions and PNe, can be considered as reliable probes of the
enrichment history of galaxies, unaffected by the immediately
preceding nucleosynthesis in the progenitor stars.  \citet{Sextans}
compared observed $\alpha$-element-to-oxygen abundance
ratios to ones for \ion{H}{ii} regions, to estimate additional
enrichment in oxygen for Type~I PN in the nearby galaxy Sextans~A.
These authors found significant self-pollution of the PN progenitor, by
a factor of $\sim$10 in oxygen.  \citet{Fornax} used the same idea
during a study of PN in the Fornax dSph galaxy and found that
systematically lower ratios for log(S/O), log(Ar/O) and log(Ne/O) in this
nebula can be easily explained with additional enrichment in
oxygen by 0.27$\pm$0.10 dex. After correction for this
additional enrichment, all studied ratios increased to the values
defined for \ion{H}{ii} regions, as shown in
Figure~\ref{fig:Abun_ratios}.  This conclusion is additionally
supported by the fact that using the same correction for the observed
log(Cl/O) ratio in the Fornax PN, moved the value to $-3.41$, consistent
with the \ion{H}{ii} regions (see panel (d) of
Figure~\ref{fig:Abun_ratios}).

\Bo\ in Sgr has a complicated abundance pattern,  which are hard to show in
Figure~\ref{fig:Abun_ratios}, since the differences are about 0.9 dex
for log(S/O), log(Ar/O) and log(Cl/O) ratios but $-0.9$ dex for
log(Ne/O).  We will try to explain neon overabundance for \Bo\ in
Section~\ref{txt:neon} below.  However, to explain the lower log(S/O),
log(Ar/O) and log(Cl/O) ratios it is natural to suggest just an
additional enrichment in oxygen, following \citet{P00} and
\citet{Sextans,Fornax}.  Using the abundance ratios for \ion{H}{ii}
regions and our observed ratios, this self-pollution can be calculated
as the weighted average, $\delta$O = 1.09$\pm$0.13 dex.  After the
correction the resulting oxygen abundance 12+log(O/H) is 6.72$\pm$0.16
dex, that is, 1/110 the solar value \citep{L03}, similar to the
metallicity of the old globular cluster Terzan 8 in Sgr \citep{DCA95}.
The corrected
ratios (shown in Figure~\ref{fig:Abun_ratios} as open circles) are
log(S/O)$_{\rm corr} = -1.54\pm0.11$, log(Ar/O)$_{\rm corr} =
-2.25\pm0.12$ and log(Cl/O)$_{\rm corr} = -3.62\pm0.23$, consistent
with the values in \ion{H}{ii} regions, showing that a change in
oxygen suffices.  Finally, we can estimate that the PN progenitor in
\Bo\ enriched the ejecta by a factor of $\sim$~12 in oxygen and by a
factor of $\sim$~240 in nitrogen.

Three of five PNe in dSph galaxies show observed log(Ne/O), log(S/O),
log(Ar/O) and log(Cl/O) ratios consistent with abundance ratios for
\ion{H}{ii} regions.  This implies that oxygen dredge-up affects
abundances only under some circumstances.  \citet{RM07} analysed
the abundances for the sample of bright PNe in dwarf irregular galaxies
and also suggested that oxygen is dredged up on occasion, even at very
low metallicity. \citet{P00} argue that oxygen is a byproduct of all
third dredge-up, but leads to enrichment only at low metallicity. At
solar metallicity, the dredged-up material has {\it lower} oxygen
abundance than the original gas.

It is uncertain why \Bo\ would show a 3$^{\rm rd}$ dredge-up (as
evidenced by its carbon-rich nature) while other PNe at similar
extreme abundances do not.  Rotational mixing might be a reason
\citep{Siess2004}.  However, if \Bo\ is a member of Sgr, it can have
a younger age and larger progenitor mass than Galactic stars of the
same metallicity, which favours the occurence of 3$^{\rm rd}$ dredge-up.

\begin{table*}
\centering{
\caption{Comparison of abundances}
\label{t:Comparison}
\begin{tabular}{lccccccccccc} \hline
\rule{0pt}{10pt}
Element  & Wray~16$-$423 & He~2$-$436 &       \St   &     \Bo    &Fornax &Solar$^a$&W$-\odot$ &He$-\odot$&     St$-\odot$&     BB$-\odot$&  F$-\odot$  \\ \hline
 He      &    11.03      &   11.03    &  11.03  &  11.00  & 10.97 & 10.99   &   +0.04  &   +0.04  &    +0.04  &    +0.09  &   $-$0.01   \\
 C       &    8.86       &   9.06     &  ~9.11  &  ~9.20  & ~9.02 & ~8.46   &   +0.40  &   +0.60  &    +0.65  &    +0.74  &     +0.56   \\
 N       &    7.68       &   7.42     &  ~7.74  &  ~7.64  & ~7.04 & ~7.90   & $-$0.22  & $-$0.48  &  $-$0.16  &  $-$0.10  &   $-$0.86   \\
 O       &    8.33       &   8.36     &  ~8.57  &  ~7.81  & ~8.01 & ~8.76   & $-$0.43  & $-$0.40  &  $-$0.19  &  $-$0.95  &   $-$0.75   \\
 Ne      &    7.55       &   7.54     &  ~7.82  &  ~7.91  & ~7.38 & ~7.95   & $-$0.40  & $-$0.41  &  $-$0.13  &  $-$0.04  &   $-$0.57   \\
 S       &    6.67       &   6.59     &  ~6.99  &  ~5.16  & ~6.45 & ~7.26   & $-$0.59  & $-$0.67  &  $-$0.27  &  $-$1.83  &   $-$0.81   \\
 Cl      &    4.89       &   ...      &  ~5.10  &  ~3.14  & ~4.60 & ~5.33   & $-$0.44  & ...      &  $-$0.23  &  $-$2.19  &   $-$0.73   \\
 Ar      &    5.95       &   5.78     &  ~6.12  &  ~4.57  & ~5.65 & ~6.62   & $-$0.67  & $-$0.84  &  $-$0.50  &  $-$2.05  &   $-$0.97   \\
 Fe      &    ...        &   ...      &  ~6.87  &  ~5.72  & ~6.38 & ~7.54   &    ...   & ...      &  $-$0.67  &  $-$1.82  &   $-$1.16   \\
\hline
\MC{12}{l}{~~} \\
\MC{12}{l}{Note: The abundances are given as 12 + log X/H.}\\
\MC{12}{l}{$^a$ Solar system abundances are from \citet{L03}.}\\
\end{tabular}
 }
\end{table*}

\subsection{Abundance Comparison}

Table~\ref{t:Comparison} shows the elemental abundance of all known
PNe in dSphs relative to the solar abundance of \citet{L03}. Values
for He\,2-436 and Wray\,16-423 are from \citet{D00}, and for the
Fornax PN from \citet{Fornax}.  \citet{D00} found that the abundances of
the first two Sgr PNe are identical within their uncertainties (0.05
dex), which provides evidence that their progenitor stars formed in
a single star burst event within a well-mixed ISM.
This star formation episode is estimated to have taken place
5 Gyr ago \citep{ZGWPHM06}. \St\ shows
significantly higher abundances and likely dates from a more recent
event of star formation.  All four objects are strongly enriched in
carbon, with C/O ratios between 3 and 29.

We find that \St\ is the most metal-rich PN known for the dwarf
spheroidal galaxies. Its [O/H] is similar to that of the few most
metal-rich PNe in the LMC \citep{Leisy2006}.  It confirms the
existence of a metal-rich population in Sgr \citep{ZGWPHM06},
in good agreement with what was found by the spectroscopic analysis
of red giants \citep{Bon04,Mo05,SBBMMZ07}. The lack
of nitrogen enrichment shows that the initial mass of the progenitor
star was $M_i < 2.5\,\rm M_\odot$ and therefore puts a lower limit to
the age of $t >1\,\rm Gyr$. The N/O versus N/H abundance ratio of \St\
closely follows the relation for Galactic PNe \citep{Leisy2006}.

\Bo\ is, in contrast, the most metal-poor object with well-determined
abundances, in Sgr.  \citet{ZGWPHM06} argue, based on limited data,
that its original abundance was [Fe/H]$<-2$. We find slightly higher
values, with S, Ar, Cl and Fe uniformly indicating an original
metallicity of [Fe/H]$=-1.9$.  The low metallicity indicates that
it belongs to the old population of Sgr
\citep[see also the discussion in][]{Be99}. The C/O$=25$ is notable,
showing that the nebular abundances are affected by extreme
self-enrichment.  In addition to C, this is also the case for N, O,
and, surprisingly, for Ne.

\subsection{The Neon Problem}
\label{txt:neon}

\Bo\ shows a very unusual abundance pattern, with neon more abundant by number
than oxygen. Oxygen and neon are both produced in Type-II supernovae, where
they are produced in a fixed ratio. The abundance ratio is Ne/O$\sim0.18$ in a
range of environments \citep{Henry1989,IT99,ISMGT06,Leisy2006}, in
confirmation of their common origin. Neon is also produced in carbon burning
high-mass AGB stars and in high-mass novae, but not in amounts exceeding
oxygen. However, significant neon can be produced in the intershell of
low-mass AGB stars, and this is the likely origin of the enrichment in \Bo. In
the large sample of \citet{Leisy2006}, the Ne/O ratio is always less than
unity: \Bo\ is unique, and either shows the result of a rather unusual event,
or traces evolution within a rare parameter range like mass and/or
metallicity.

The intershell contains the ashes of the preceding hydrogen burning.
Following helium ignition, this intershell becomes convective and experiences
burning at $T\sim 3 \times 10^8$\,K \citep{Herwig2005}. The nitrogen is
burned to $^{22}$Ne via two $\alpha$ captures. In high mass stars, hot bottom
burning destroys the neon, but in low-mass stars this does not occur
and neon increases to $2\% $ by mass \citep{Herwig2005}.

Intershell material is usually exposed only in hydrogen-poor stars,
which are likely to have lost their hydrogen layer in a very late thermal
pulse. These show significant enhancement in Ne \citep{Werner2004},
in two cases also with low O/Ne.
Interestingly, the two cases with low O/Ne also show
surface hydrogen, although still with H/He$\, \sim 1$ by mass. 

\Bo\ shows no evidence for hydrogen depletion, or of a separate
hydrogen-poor region, and is therefore probably not related to the
hydrogen-poor stars. Its intershell products instead are visible due to
dredge-up.  The neon abundance in \Bo\ is about 0.15\%\ by mass,
compared to 2\%\ expected in the intershell. This suggests dilution by
a factor of 15. Assuming an intershell mass of $\sim
10^{-2}\,$M$_\odot$ \citep{Herwig2005} we find that the envelope mass
at the time of the mixing would have been $\sim 0.1\,\rm M_\odot$.
Assuming an initial metallicity of the envelope of 0.01\,Z$_\odot$, we
find that the final surface abundances of C, N, O, Ne are dominated by
the intershell products, while the helium abundance remains dominated
by the original envelope. This can consistently explain the observed
abundances.  The C/O ratio in the intershell is $\sim 20$
\citep{Iben95}, which approaches the value seen in \Bo.

\section{Conclusions}
\label{txt:summ}

The first scientific results based on SALT observations were published
by \citet{Dono06,W06} and \citet{BK07}, based on imaging data.  In
this paper we present the first spectrophotometric results obtained
with SALT and RSS during their performance verification phase that
emphasize the long-slit capabilities of the RSS for spectrophotometric
observations.  The quality of our data permits us to measure line
ratios of elements not accessible in earlier studies.  We measured the
electron temperatures, the electron densities and element abundances
for O, N, Ne, Ar, S, Cl, Fe, C and He elements in two PNe in the
Sagittarius dSph galaxy.  These results are presented in several
tables and plots.  Based on the data and discussion presented in the
paper, the following conclusions can be drawn:

1. We confirm that \St\ is the most metal-rich PN known in any dwarf
   spheroidal galaxies and has an oxygen abundance of 12+log(O/H) =
   8.57$\pm$0.02 dex.  This [O/H] shows that Sgr contains a younger
   stellar population with [Fe/H]$\approx-0.2$, in good agreement with
   spectroscopic abundance measurements in Sgr stars. The element abundance
   ratios of Ne, S, Ar and Cl to O for \St\ are  consistent with
   the expected patterns in \ion{H}{ii} regions showing absence of any
   additional enrichment in oxygen in the PN progenitor.  The lack of
   nitrogen enrichment puts a lower limit to the age of PN progenitor
   $t >1\,\rm Gyr$.

2. We obtain an oxygen abundance of 12+log(O/H) = 7.81$\pm$0.09 for
   \Bo.  According to our analysis, this value should be corrected
   downward by 1.09$\pm$0.13 dex due to the self-pollution of oxygen
   by the PNe progenitor. After this correction the element abundance
   ratios S/O, Ar/O and Cl/O appear in a good overall accord with the
   trends seen for H\,{\sc ii} regions.  This implies that the PN
   progenitor had an oxygen abundance of 12+log(O/H) =
   6.72$\pm$0.16 dex, or 1/110 of the solar value, similar to the
   metallicity of the old globular cluster Terzan 8 in Sgr \citep{DCA95}.
   \Bo\ is therefore one of the most metal-poor PNe with well-determined
   abundances and belongs to the old population of Sgr.
   During its normal stellar evolution, the PN progenitor enriched
   the material by a factor of $\sim$~12 in
   oxygen and by a factor of $\sim$~240 in nitrogen.

3. \Bo\ shows a very unusual pattern with neon being more abundant by number
   than oxygen.  This neon enrichment could be explained by production
   of neon in the intershell of low-mass AGB stars. In this case the
   final surface abundances of C, N, O, Ne are dominated by the
   intershell products, that can consistently explain the observed
   abundances.

4. We confirm the existence of Wolf-Rayet broad lines in \St, but no
   WR features were detected in the spectrum of \Bo.  The fraction of
   [WR]-type central PNe stars is 67\% for dSph galaxies, that
   supports models in which the probability of star to develop a WR wind
   is determined by parameters of the progenitor star.

\section*{Acknowledgments}

We thank the anonymous referee for comments which improved the
presentation of the manuscript.
This paper was written while AAZ was a sabbatical visitor at the
South African Astronomical Observatory in Cape Town; AAZ is
grateful for the hospitality.
We are grateful for the support of numerous people during the
SALT  PV phase.
L.S.P. acknowledges the partial support of the Cosmomicrophysics program
of the National Academy of Sciences and Space Agency of Ukraine.


\bsp

\label{lastpage}

\end{document}